\documentclass[prl,twocolumn,showpacs]{revtex4}
\usepackage{natbib}
\usepackage{graphicx} 

\begin{document}

\title
{Sodium-terminated zigzag graphene nanoribbon: A one-dimensional
semimetal with a tilted Dirac cone}

\author{Ozan Ar\i}
\affiliation{Department of Physics, Izmir Institute of Technology,
35430 Izmir, Turkey}

\author{\.{I}brahim Dursun}
\affiliation{Department of Physics, Izmir Institute of Technology,
35430 Izmir, Turkey}

\author{R. Tu\u{g}rul Senger}
\email{tugrulsenger@iyte.edu.tr}
\affiliation{Department of Physics, Izmir Institute of Technology, 35430 Izmir,
Turkey}

\date{\today}

\begin{abstract}

Using first principles pseudopotential density functional theory
calculations, we find that terminating zigzag graphene nanoribbons
(ZGNR) with monovalent alkali atoms at a reduced concentration has
a dramatic impact on their properties. In particular, using sodium
atoms for the saturation of ZGNR edges at half the concentration
of edge-carbon atoms make it a  one dimensional, perfect
semimetal, where the valance and conduction bands meet at only a
single, Dirac-like point. Unlike pristine graphene, the
Dirac-``cones'' of Na-ZGNR is not symmetric with respect to wave
vector, but rather it is tilted.

\end{abstract}

\maketitle
After its realization as a stable monolayer structure
\cite{novo1}, graphene has become a material of great interest for
research in material science due to its impressive physical
properties and the potential to tune them
\cite{novo2,kat,zhou,geim,novo3}. Pristine graphene is a
zero-bandgap (perfect) semimetal. The conduction and valance bands
of graphene meet linearly at six distinct points (Dirac-points) of
first Brillouin zone, and are in the form of symmetric cones at
the Fermi level \cite{neto2009electronic}. Long and thin strips of
graphene, known as graphene nanoribbons (GNR), are semiconducting
and they have been experimentally realized with smooth edges and a
range of widths down to sub 10 nm scale
\cite{tapa,li2008chemically}. As a special class, zigzag-edged
GNRs (ZGNR) have width dependent energy band gaps, and are also
expected to present edge localized spin states
\cite{yama,lee,pisani,kusakabe,barone}. These localized spin
states are ferromagnetically ordered at each edge and they are
antiferromagnetically coupled to each other
\cite{fujita1996peculiar}.

In computational studies of ZGNRs, to facilitate mechanical
stability of the structure, the edges are usually terminated with
hydrogen atoms, which lowers the values of its energy band gap and
the induced magnetic moments of the edges. To manipulate the
electronic properties of ZGNRs, modification \cite{gunlycke2007altering,jiang2007unique} and functionalization \cite{boukhvalov2008chemical} of edges and
producing defects, doping \cite{martins2007electronic,martins2008,yu2008first,zheng2009effects} and absorbtion
\cite{rigo2009electronic,gorji,ataca2011adsorption} of atoms or
molecules on ZGNRs have been extensively studied. For instance, recent
studies \cite{kan2008half,son2006half} have shown ways to
obtain half-metallic ZGNRs.

In this letter we report that edge saturation of ZGNRs with Na
atoms at a specific concentration makes them a zero-gap one
dimensional semimetal with an asymmetric linear energy dispersion
around the Fermi level. That corresponds to a Dirac cone similar
to pristine two dimensional (2D) graphene
\cite{neto2009electronic}, however, with a tilted form. Similar to
hydrogenization of the edges, Na atoms saturates the dangling
bonds of edge carbon atoms, and modify the local magnetic moment
magnitudes and change the electronic band structure.

Geometry optimizations of bare and Na-terminated ZGNRs and
calculations of their magnetic and electric properties were
performed using the SIESTA package \cite{siesta} based on density
functional theory (DFT) \cite{khon}. We have used generalized
gradient approximation (GGA) for the exchange and correlation
potential as parameterized by Perdew, Burke and Ernzerhof
\cite{perdew}. For geometry optimizations, a local relaxation has
been performed using the conjugate gradient algorithm and the
convergence criteria of 0.04 eV$\times$\AA$^{-1}$ and $10^{-4}$ eV for
the forces and total energies, respectively, in the
self-consistency cycles. The electrostatic potentials were
determined on a real-space grid with a mesh cutoff energy of 300
Ry. We make use of non-conserving Troullier Martins
pseudopotentials \cite{troul} in the Kleinman Bylander factorized
form \cite{klein} and a double-$\zeta$ polarized basis set composed of
numerical atomic orbitals of finite range. The Brillouin zone has
been sampled with (1,1,70) points within the Monkhorst-Pack
k-point sampling scheme.

 \begin{figure}
 \includegraphics[width=4.5 cm]{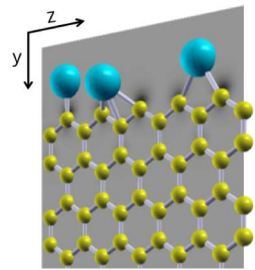}
 \caption{(Color online) Schematic representation of various
 binding geometries of Na atoms to a ZGNR edge. From left to right:
 top in-plane (TI), bridge off-plane (BO), bridge in-plane
 (BI).}\label{geo}
 \end{figure}

\begin{figure}
 \includegraphics[width=8.5 cm]{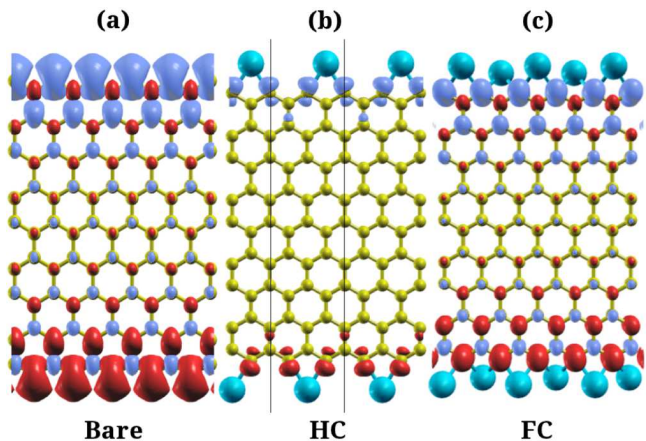}
 \caption{(Color online) Relaxed geometric structure and
 isosurfaces of charge density difference of spin-up ($\uparrow$)
 and spin-down ($\downarrow$) states for 9-ZGNR; bare (a), and Na
 terminated at HC (b) and FC (c) concentrations. For all geometries, positive and negative
 values of the charge density difference is shown by red (dark) and
 blue (light) regions, respectively, for the same isosurface value
 of $\pm 0.0025$ electrons/\AA$^3$.
 }\label{charge}
 \end{figure}

We have considered edge termination of $n$-ZGNRs ($n$ denoting the
number of zigzag rows in the ribbon) with monovalent atoms, i.e.
H, Li, Na and K. Here we will report our results using Na
terminated 9-ZGNR as a representative case. In Fig.\ref{geo},
different binding geometries of Na atoms to ZGNR determined after
geometry optimizations are shown; these are, on top of an edge C
atom, in-plane with the ribbon (TI); bridge site of two edge C
atoms, either off-plane (BO), or in plane (BI). Unlike
H-termination, Na atoms prefer to bind at the bridge sites rather
than the TI site; binding energy of a single Na atom at a BI site
is 0.46 eV larger than that at a TI site. We considered two
different concentrations of Na for saturation of the edges, full
coverage (FC) where there are as many Na atoms as the edge C
atoms, and half coverage (HC) where the Na atoms are bound to
every other bridge site only (see Fig.\ref{charge} for the unit
cell of the ribbon and Na binding sites). In the HC case,
equilibrium position of Na atoms is BI with a Na-C bond length of
2.33 \AA. In the FC case, however, due to their larger atomic size
than H, neighboring Na atoms relax towards opposite directions at
the BO sites. In this case, Na-C bond length and vertical distance
of Na to the plane of ribbon are 2.35 \AA \, and 1.23 \AA \,
respectively. Accordingly, the binding energy per Na atom reduces
to 2.84 eV at FC from 2.91 eV at HC.

We calculated spin-dependent charge densities of the optimized
structures and their difference ($\Delta \rho =
\rho_{\uparrow}-\rho_{\downarrow})$. The characteristic
edge-localized spin states of ZGNRs are preserved for both HC and
FC concentrations of Na terminated 9-ZGNR (see Fig.\ref{charge}).
Using Mulliken population analysis we have determined spin
dependent atomic charges, total valance charge and the spin
magnetic moments of the atoms. Total magnetic moments of the
ribbons are found zero for all cases. As seen in the figure, there
are no net induced magnetic moments on the Na atoms either. The
penetration of the spin imbalance on the C atoms, $\Delta \rho$,
toward the center of ribbons is considerably shorter for the HC
case. Binding of Na atoms reduces the magnetic moments of the edge
carbon atoms from $\mu = 1.21 \, \mu_B$ to $\mu = 0.11 \, \mu_B$
for HC, and to $\mu = 0.28 \, \mu_B$ for FC. This magnetic moment
reduction is comparable to the full hydrogenation of ZGNRs where
edge carbon magnetic moments were reported as $\mu = 0.26 \,
\mu_B$ \cite{sahin2008first}. When a Na atom binds to ZGNR edge,
its valance charge is partially transferred to the ZGNR due to
lower electronegativity of Na than C. At HC (FC) concentration, Na
transfers 10\% (15\%) of its total valance charge.
Correspondingly, total valance charge of each edge carbon atom of
9-ZGNR increases from 4.06 for bare ZGNR to 4.13 at HC and 4.20 at
FC concentration.

\begin{figure}
\includegraphics[width=8.5cm]{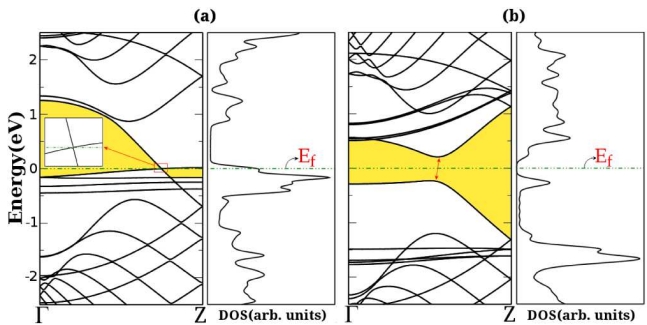}
\caption{(Color online) Band structures and density of states
diagrams for 9-ZGNR. (a) HC, and (b) FC concentration of Na. The
inset magnifies the tilted Dirac cone at the Fermi level for the
HC case.}\label{band}
\end{figure}

In Fig.\,\ref{band} we display the energy band diagram and density
of states of Na terminated 9-ZGNR at HC and FC concentrations.
Each case modifies the electronic properties of the ribbon
differently. We calculate that bare 9-ZGNR is a direct gap
semiconductor with $E_g = 0.70$ eV. Fully hydrogenated 9-ZGNR has
a lower band gap of $E_g = 0.50$ eV, consistent with a previous
report \cite{son2006energy}. Similarly, 9-ZGNR terminated with Na
at FC concentration has a reduced and slightly indirect band gap
of $E_g = 0.44$ eV.

At HC concentration of Na termination the 9-ZGNR becomes a
zero-gap semimetal with an interesting band structure. The valance
and conduction bands intersect at a single point. In
Fig.\,\ref{band}(a), the Dirac-like point is located at $k_0 =
0.469 \,$ \AA$^{-1}$ in the first Brillouin zone ($k_Z=0.625$
\AA$^{-1}$). The tilted Dirac cone is formed by a steep linear
band crossing a relatively flat band at the Fermi level. These
bands are mainly derived from carbon $p_x$ and combination of
carbon $p_y$ and Na $s$ orbitals, respectively. Since the present
ribbon structure is one dimensional, the density of states (DOS)
at the Fermi level does not vanish unlike 2D-graphene, and also,
singularities characteristic to one-dimensional systems is evident
in the DOS plot.

Energy band structures showing Dirac cones are quite rare.
2D-graphene is known for having symmetric and isotropic Dirac
cones at corners of the first Brillouin zone. In this context,
recently another two dimensional structure, which is an organic
compound, has been reported as a unique material having a gapless
band diagram and asymmetrical linear energy dispersion
\cite{goerbig2008tilted}. Na terminated ZGNR at HC concentration
that we present in this study is a one dimensional member of this
class of materials having tilted Dirac ``cones''.

Around the Fermi level, $E$ vs $k$ relation can be approximated in the form,
\begin{eqnarray}
 E_{\lambda} (k) = w_0 \, (k - k_0) + \lambda \, w \, |k-k_0|
\end{eqnarray}
where $\lambda$ plays the role of band index ($+1$ for conduction, and $-1$ for valance band), $w=2.517$
eV$\times$\AA\, and $w_0=-2.277$ eV$\times$\AA\,  are effective velocities.
In this representation the $w_0$ parameter determines the degree of tilting of the Dirac cone.
The slopes of the bands at the Fermi level are $w_0+w= 0.240$ eV$\times$\AA \, and
$w_0-w=4.794$ eV$\times$\AA\,, so that an asymmetry of about a factor of 20 in the slopes of the fermion and
antifermion bands is present.

We find that the ``tilted Dirac-cone'' shaped band structure is
quite robust with respect to the ribbon width and the
approximations involved, such as exchange correlation functional
being LDA or GGA. Using Li or K instead of Na also produces
similar electronic structures, however, the valance and conduction
bands intersect at slightly below or above the Fermi energies,
respectively.

In summary, we have performed first-principles total energy
calculations to investigate edge saturation of ZGNRs with Na
atoms. Binding a Na atom to every other bridge site of the zigzag
edge leads to formation of linear bands crossing at the Fermi
level. To our knowledge, such tilted Dirac ``cones'' in a one
dimensional material has not been reported before
\cite{X.L.Wang2010}, and may find interesting spintronics
applications. Such a one-dimensional semi-metal with charge
carriers having high mobility is desirable for realization of
nanoscale devices. Our preliminary results also show that edge
saturation of ZGNRs with other monovalent alkali metals (Li and K)
have similar effects, but those will be reported elsewhere.

\begin{acknowledgements}
This work was partially supported by \.IzTech-BAP program through
project no: 2010\.IYTE26.
\end{acknowledgements}

\end{document}